\documentclass[twocolumn,a4paper]{article}
\usepackage{epsfig}% SRC Specials
\newcommand{\be}{\begin{equation}}
\newcommand{\ee}{\end{equation}}
\newcommand{\ba}{\begin{eqnarray}}
\newcommand{\ea}{\end{eqnarray}}
\newcommand{\ban}{\begin{eqnarray*}}
\newcommand{\ean}{\end{eqnarray*}}
\begin{document}

\title{\Large\bf Is there a real time ordering behind the nonlocal correlations?}

\author{Antoine Suarez\thanks{suarez@leman.ch}\\{\it\small Center for Quantum Philosophy, P.O. Box 304, CH-8044
Zurich, Switzerland}}

\date{October 20, 2001}

\maketitle

{\footnotesize \emph{Abstract}: It is argued that recent
experiments with moving beam-splitters demonstrate that there is
no real time ordering behind the nonlocal correlations: In Bell's
world there is no ``before'' and ``after''.}

{\footnotesize PACS numbers : 03.65.Bz, 03.30.+p, 03.67.Hk,
42.79.J} \vspace{2mm}

Quantum Mechanics predicts correlated outcomes in space-like
separated regions for experiments using two-particle entangled
states  \cite{jb64}. But two events cannot be correlated if each
of them takes place quite at random: ``Correlations cry out for
explanation'' \cite{jb64}, either the events are pre-determined by
some kind of hidden variables, or directly connected through some
kind of influence at a distance.

If all influences in nature stick to time and never pace faster
than light, the quantum correlations in space-like separated
regions imply particles carrying hidden variables, which determine
the particle's behaviour. Apparently this was the way Einstein
thought, and concluded that the quantum mechanical description of
the physical reality cannot be considered complete \cite{epr}.
However Bell showed that if one only admits relativistic local
causality (causal links with $v\leq c $), the correlations
occurring in two-particle experiments should fulfill clear
locality conditions (``Bell's inequalities'') \cite{jb64}. Bell
experiments conducted in the past two decades, in spite of their
loopholes, suggest a violation of local causality: statistical
correlations are found in space-like separated detections;
violation of Bell's inequalities ensure that these correlations
are not pre-determined by local hidden variables \cite{exp}.
Nature seems to behave nonlocally, and Quantum Mechanics predicts
well the observed distributions.

If the correlations are not pre-determined, then there are
influences connecting directly two space-like separated events,
even though we cannot use such ``Bell influences'' for
faster-than-light communication \cite{eb78}. Suppose one of the
measurements produces the value $\rho$ ($\rho\in\{+,-\}$), and the
other the value $\sigma$ ($\sigma\in\{+,-\}$). Quantum Mechanics
predicts a probability $Pr(\rho, \sigma)$ of getting the outcome
$(\rho,\sigma)$, which is independent of any ordering of the
events. Nevertheless this does not mean that Quantum Mechanics is
incompatible with the whole idea of ordering. Effectively, one can
consider that the value $\sigma$ depends on the value $\rho$ (what
is not the same as saying that $\sigma$ is caused by $\rho$), and
the correlations are worked out using the conditional probability:

\begin{footnotesize}
\ba Pr(\sigma|\rho)=\frac  {Pr(\rho, \sigma)}{\sum_{\sigma}
Pr(\rho, \sigma)} \label{Pr12}\ea
\end{footnotesize}

\noindent where $Pr(\sigma|\rho)$ is the probability that one of
the particles produces the value $\sigma$ provided the other
produces the value $\rho$, and $Pr(\rho, \sigma)$ is the quantum
mechanical joint probability of getting the outcome
$(\rho,\sigma)$.

But one can, alternatively, consider that $\rho$ depends on
$\sigma$ and the correlations are worked out using the conditional
probability:

\begin{footnotesize}
\ba Pr(\rho|\sigma)=\frac{Pr(\rho, \sigma)}{\sum_{\rho} Pr(\rho,
\sigma)} \label{Pr21} \ea
\end{footnotesize}

In spite of the different orderings the equations (\ref{Pr12}) and
(\ref{Pr21}) bear the same joint probability of getting the
outcome $(\rho,\sigma)$, i.e., the quantity $Pr(\rho, \sigma)$
predicted by Quantum Mechanics, which is ordering independent. To
produce the correlations, nature can (arbitrarily) choose between
the ordering assumed in (\ref{Pr12}) and that in (\ref{Pr21}) but
its choice has no observable consequence at all, and it is not
possible, even in principle, to distinguish which measurement is
the independent and which the dependent one. So what Quantum
Mechanics actually implies is that the order the nonlocal
correlations reveal does not correspond to any real time ordering
and, consequently, is not tied to any experimentally
distinguishable frame. Suppose a physicist could act nonlocally
and would like to bring about Bell-correlations: he should choose
one event as first, assign it a value at random, and then assign a
value to the other event depending on the value assigned to the
first. Quantum Mechanics actually means that in nature this
ordering activity comes about without flow of time.

Of course, in case of experiments with time-like separated
measurements \cite{ptjr94} one has to accept that the measurement
occurring later in time takes account of that occurring before:
indeed in this situation one could even arrange that the outcome
in side 1 of the setup determines classically (through light
signals) the phase in side 2 and, thereby, the outcome in side 2.

The time-independence of Quantum Mechanics is probably its most
astonishing feature. Indeed we use to explain correlations in the
physical world assuming that a ``temporally" earlier event
influences a ``temporally" later event. As long as one believed
(following Einstein) that there are no space-like influences, the
fundamental temporal notion could not be other than proper time
along a time-like trajectory. But since Bell experiments did
reveal us a world consisting in nonlocal influences, the
``reasonable'' position in the very spirit of the relativity of
time is to assume that these influences can be described using
several simultaneity lines to distinguish between ``before'' and
``after''. Therefore, also taking nonlocality for granted, the
decisive question remains: is there an experimentally
distinguishable time ordering behind the nonlocal correlations?

The first attempt to cast nonlocality into a temporal scheme has
been Bohm's theory \cite{dbbh}. It uses a unique preferred frame
or absolute time, in which one event is caused by some earlier
event by means of instantaneous action at a distance. This
description makes the same predictions as Quantum Mechanics, and
the assumed instantaneous influences cannot be used for
superluminal communication. The assumption of one preferred frame
has been invoked recently as the most natural way to incorporate
quantum nonlocality \cite{lh92}. Nevertheless,  if one tries to
cast nonlocal causality into only one preferred frame it is not
more reasonable to connect a "cause" event to an "effect" event in
that frame rather than in some other frame. Effectively a single
preferred frame (``quantum ether'') is ``experimentally
indistinguishable" \cite{jb64}: The predictions would remain the
same if one assumes that the preferred frame is a virtual entity
changing from experiment to experiment. One is tempted to think
that Bohm introduces absolute time just because he wishes to
justify a causal description, but in the end, an untraceable
``quantum ether'' is essentially the same as deciding arbitrarily
which event depends on which one. What is more, in the particular
case, possible in principle, of both measurements taking place at
exactly the same time in the preferred frame, the only way of
establishing which event depends on which is by arbitrary
decision. Actually Bohmian Mechanics and any theory using only one
preferred frame, can be considered a causal description but not a
real temporal one.

Work in recent years proposed to imbed nonlocality in a real
relativistic time ordering by using several relevant frames. The
main motivation of such a proposal is to create an experimental
test allowing us to decide whether nonlocal influences can be
measured by means of several real clocks. The result is a nonlocal
description called Multisimultaneity or Relativistic Nonlocality
\cite{asvs97,as97,as00.1}. Aiming an explanation in terms of
several real clocks, it is natural to assume that these clocks are
somewhat related to the frames involved in the experimental setup.
More specifically these frames are supposed to be those of the
devices in which the choice of the outcome value occurs, and the
monitored beam-splitters are supposed to be these
``choice-devices'' \cite{ophoc}. The basic assumption of
Multisimultaneity is that the decision about the output port by
which a photon leaves a beam-splitter takes account of all the
local and nonlocal information available within the inertial frame
of this choice-device, at the instant the particle strikes it; we
stress that this frame is unambiguously defined by the velocity
corresponding to the Doppler-shift of the reflected photons
\cite{as00.1}. Within each choice-device's frame the causal links
always follow a well defined chronology, one event never depending
on some future event.

In the conventional Bell experiments both beam-splitters are
standing still in the laboratory frame. In this frame one of the
choices (say $\rho$) takes place always before the other
($\sigma$), and the particle arriving later takes account of the
decision of that arriving first, just as indicated in equation
(\ref{Pr12}). Therefore, Multisimultaneity bears the same
predictions as Quantum Mechanics. In this sense, Multisimultaneity
and Bohm's theory provide basically the same description for
experiments with all choice-devices at rest, i.e. a causal
explanation in which the ordering of the events fits with the time
ordering in the laboratory frame.

Consider now experiments in which the choice-devices are in motion
in such a way that each of them, in its own reference frame, is
first to select the output of the photons (\emph{before-before}
timing). Then, each particle's choice will become independent of
the other's, and according to Multisimultaneity the nonlocal
correlations should disappear. By contrast Quantum Mechanics
requires that the particles stay nonlocal correlated independently
of any timing, even in such a \emph{before-before} situation
\cite{as00.1}. This means that before-before experiments are
capable of acting as standard of time-ordered nonlocality: if
Quantum Mechanics prevails, nonlocality cannot be imbedded in a
relativistic chronology; if Quantum Mechanics fails, there is a
time ordering behind the nonlocal correlations, and proper time
along a time-like trajectory is not the only temporal notion
\cite{ip98}.

In summary, Multisimultaneity proposes feasible experiments using
moving choice-devices, which are of interest in the general
context of physical situations involving several observers in
relative motion \cite{pe01}. Acousto-optic modulators have made it
possible to perform such experiments \cite{as00.1}. The results
recently obtained with before-after, before-before, and
after-before timings uphold the predictions of Quantum Mechanics
\cite{szsg}. This means that in before-before experiments the
correlations are caused regardless of any relativistic chronology:
entangled photons run afoul of the relativity of time.

%%%%%%%%%%%%%%%%%%%%%%%%%%%%%%
\begin{figure}[t]
\centering\epsfig{figure=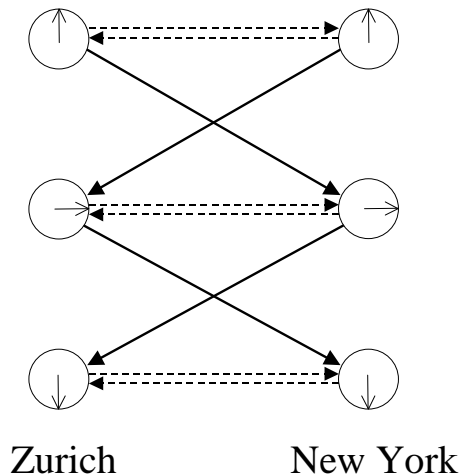,width=60mm}
{\caption{\footnotesize The influences allowing us to phone
between two separated regions follow time-like trajectories
(solid), and can consistently be described in terms of ``before"
and ``after" by means of real clocks; Einstein's world contains
only such local causal links. The influences bringing about
nonlocal correlations (dash) cannot be described in terms of
``before" and ``after" by means of any set of real clocks. Bell's
world consists in such nonlocal influences. The notion of time
makes sense only in Einstein's world, i.e. along time-like
trajectories.}} \label{Seriesfig1}
\end{figure}
%%%%%%%%%%%%%%%%%%%%%%%%%%%%%%

The influences allowing us to phone between two separated regions
follow time-like trajectories (Fig. 1), and can consistently be
described in terms of ``before" and ``after" by means of real
clocks; Einstein's world contains only such local causal links.
The influences bringing about nonlocal correlations cannot be
described in terms of ``before" and ``after" by means of any set
of real clocks. Bell's world consists in such nonlocal influences.
The notion of time makes sense only in Einstein's world, i.e.
along time-like trajectories.

Taken for granted the impossibility of imbedding nonlocality in a
relativistic chronology, we would like to finish our analysis by
discussing the possibility of imbedding it in a non-relativistic
temporal scheme that uses influences propagating at finite
velocity $V>c$. Such a theory has been proposed by Eberhard
\cite{eb89}. The value of the new constant $V$ is not given, but
possible experiments are described, which would allow us to
establish it providing they prove standard Quantum Mechanics
wrong. Since the preferred frame is in principle experimentally
distinguishable, it would define a real universal clock and,
therefore, the assumed causality is a temporal one. However, the
experiments proposed would not be capable of discarding the
preferred-frame description: upholding of the quantum mechanical
predictions would simply establish a lower bound for the speed $V$
of the superluminal influences causing the correlations. As
Eberhard himself shows, the influences propagating at a finite
speed $V>c$ could be used for superluminal communication
\cite{eb89}. Although such signaling contradicts in every sense
the convictions of what is feasible in physics after
Michelson-Morley and related experiments, it doesn't bear causal
loops while only one single frame is assumed. Thus, the belief in
a \emph{real} preferred frame, if it is serious, should lead to
\emph{real} experiments aiming to demonstrate superluminal
signaling \cite{ophoc}. But more than signaling, we fear it is the
fact that the theory cannot be falsified when tested against
Quantum Mechanics that prevents physicists from performing the
proposed experiments.

In conclusion, Multisimultaneity was the product of two things:
the fact that nonlocal correlations evidence an ordering activity
(they cannot just result by random choice), and the bias that such
an ordering is always tied to a corresponding distinguishable
temporal sequence \cite{co}. Experiments with moving
beam-splitters indicate that the nonlocal correlations are brought
about without relation to any real chronology: we have to give up
the bias and accept time-independent Bell influences. Bell showed
that Einstein's reality is not the whole physical reality, and
discovered us a world without distances. The experimental results
we have discussed strong suggest that in Bell's world things get
along without any real time either.

I would like to thank Nicolas Gisin, Valerio Scarani, Andr\'e
Stefanov, and Hugo Zbinden for very stimulating discussions, and
the Odier Foundation of Psycho-physics and the L\'eman Foundation
for support.

\begin{footnotesize}

\end{footnotesize}

\end{document}